\documentclass[twocolumn,showpacs,preprintnumbers,amsmath,amssymb,floatfix]{revtex4}

\usepackage{graphicx}
\usepackage{dcolumn}
\usepackage{bm}

\begin{document}

%

\let\a=\alpha      \let\b=\beta       \let\c=\chi        \let\d=\delta
\let\e=\varepsilon \let\f=\varphi     \let\g=\gamma      \let\h=\eta
\let\k=\kappa      \let\l=\lambda     \let\m=\mu
\let\o=\omega      \let\r=\varrho     \let\s=\sigma
\let\t=\tau        \let\th=\vartheta  \let\y=\upsilon    \let\x=\xi
\let\z=\zeta       \let\io=\iota      \let\vp=\varpi     \let\ro=\rho
\let\ph=\phi       \let\ep=\epsilon   \let\te=\theta
\let\n=\nu
\let\D=\Delta   \let\F=\Phi    \let\G=\Gamma  \let\L=\Lambda
\let\O=\Omega   \let\P=\Pi     \let\Ps=\Psi   \let\Si=\Sigma
\let\Th=\Theta  \let\X=\Xi     \let\Y=\Upsilon

%

%

\def\cA{{\cal A}}                \def\cB{{\cal B}}
\def\cC{{\cal C}}                \def\cD{{\cal D}}
\def\cE{{\cal E}}                \def\cF{{\cal F}}
\def\cG{{\cal G}}                \def\cH{{\cal H}}
\def\cI{{\cal I}}                \def\cJ{{\cal J}}
\def\cK{{\cal K}}                \def\cL{{\cal L}}
\def\cM{{\cal M}}                \def\cN{{\cal N}}
\def\cO{{\cal O}}                \def\cP{{\cal P}}
\def\cQ{{\cal Q}}                \def\cR{{\cal R}}
\def\cS{{\cal S}}                \def\cT{{\cal T}}
\def\cU{{\cal U}}                \def\cV{{\cal V}}
\def\cW{{\cal W}}                \def\cX{{\cal X}}
\def\cY{{\cal Y}}                \def\cZ{{\cal Z}}
%

\newcommand{\Ns}{N\hspace{-4.7mm}\not\hspace{2.7mm}}
\newcommand{\qs}{q\hspace{-3.7mm}\not\hspace{3.4mm}}
\newcommand{\ps}{p\hspace{-3.3mm}\not\hspace{1.2mm}}
\newcommand{\ks}{k\hspace{-3.3mm}\not\hspace{1.2mm}}
\newcommand{\des}{\partial\hspace{-4.mm}\not\hspace{2.5mm}}
\newcommand{\desco}{D\hspace{-4mm}\not\hspace{2mm}}



\title{\boldmath 
 $B\to K^*\ell^+\ell^-$ Forward-backward Asymmetry and New Physics }
\author{Artyom Hovhannisyan$^{a,}$\footnote{On leave from
                                   Yerevan Physics Institute,
                                   Yerevan, Armenia.}
                                           }
\author{Wei-Shu Hou$^{b}$}
\author{Namit Mahajan$^b$}
\affiliation{
 $^a$Center for Theoretical Sciences, National Taiwan
 University, Taipei, Taiwan 10617, R.O.C. \\
 $^b$Department of Physics, National Taiwan
 University, Taipei, Taiwan 10617, R.O.C.
}
\date{\today}

\begin{abstract}
The forward-backward asymmetry ${\cal A}_{\rm FB}$ in $B\to
K^*\ell^+\ell^-$ decay is a sensitive probe of New Physics.
Previous studies have focused on the sensitivity in the position
of the zero. However, the short distance effective couplings are
in principle complex, as illustrated by $B\to \rho\ell^+\ell^-$
decay within the Standard Model. Allowing the effective couplings
to be complex, but keeping the $B\to K^*\gamma$ and
$K^*\ell^+\ell^-$ rate constraints, we find the landscape for
${\cal A}_{\rm FB}(B\to K^*\ell^+\ell^-)$ to be far richer than
from entertaining just sign flips, which can be explored by future
high statistics experiments.
\end{abstract}

\pacs{
 13.25.Hw, 
 13.20.-v, 
 12.60.-i 
 }
\maketitle


It was pointed out 20 years ago~\cite{HWS} that the loop-induced
$bsZ$ coupling is enhanced by large $m_t$, which turns out to
dominate $b\to s\ell^+\ell^-$ ($\bar B \to X_s\ell^+\ell^-$)
decay. The effective $bs\gamma$ coupling gives a low $q^2 \equiv
m_{\ell\ell}^2$ peak in the differential rate~\cite{Houmumu},
while $Z$ and $\gamma$ induced amplitudes interfere across the
$q^2$ spectrum. One such effect is the forward-backward
asymmetry~\cite{AFB}, ${\cal A}_{\rm FB}$, which is the asymmetry
between forward and backward moving $\ell^+$ versus the $B$ meson
direction in the $\ell^+\ell^-$ frame.


The first measurement of ${\cal A}_{\rm FB}$ in exclusive $B\to
K^*\ell^+\ell^-$ decay was recently reported~\cite{AFBBelle} by
the Belle experiment, with 3.4$\,\sigma$ significance. The results
are consistent with the Standard Model (SM), rules out the wrong
handed $\ell^+\ell^-$ current, but a sign flip of the $bs\gamma$
coupling is still tolerated by the poor statistics of $\sim 100$
signal events. However,
taking the measured inclusive $b\to s\gamma$ ($\bar B \to
X_s\gamma$) and $b\to s\ell^+\ell^-$ rates together~\cite{GHM},
the
latter possibility is disfavored.

The relative insensitivity of ${\cal A}_{\rm FB}$ to hadronic
effects makes it an attractive probe for New Physics (NP) in the
long run. For example, we expect a quantum jump in the number of
events with the advent of LHC in 2008. A study by the LHCb
experiment shows that $\sim$ 7700 $B\to K^*\ell^+\ell^-$ events
are expected with 2~fb$^{-1}$ data~\cite{LHCb2fb}. In this Letter
we point out that the sensitivity of ${\cal A}_{\rm FB}$ to NP is
greater than previously thought. 
The {\it complexity} of the associated effective Wilson
coefficients can be probed by $d{\cal A}_{\rm FB}/dq^2$ as early
as 2008 at the LHC.


The quark level decay amplitude is~\cite{HWS,ABHH},
\begin{eqnarray}
 {\cal M}_{b\to s\ell^+\ell^-}
 & = & -\frac{G_F \alpha}{\sqrt{2} \pi} \, V_{cs}^\ast V_{cb} \,
       \left\{
        C_9^{\rm eff}\, [\bar{s} \g_\mu L b] \,
                        [\bar\ell\g^\mu\ell] \right. \nonumber \\
 & &  \ \ \ \ \ \ \ \ \ \ \;+\; C_{10}\; [\bar{s} \g_\mu L b] \,
                   [\bar\ell\g^\mu\gamma_5\ell] \nonumber \\
 & &
      -\; 2 \frac{\hat{m}_b}{\hat s}\, C_7^{\rm eff} \left.
               [\bar{s}\, i\sigma_{\mu\nu}\hat q^\nu R b] \,
               [\bar\ell\g^\mu\ell]\right\},
        \label{eq:Amp}
\end{eqnarray}
where 
$s = q^2$, and we normalize by $m_B$, e.g. $\hat s=s/m_B^2$. We
factor out $V_{cs}^\ast V_{cb}$ instead of the usual $V_{ts}^\ast
V_{tb}$. Although trivial
within SM, it has the advantage of being real and in terms of CKM
elements that are already measured. Short distance physics,
including within SM, are isolated in the Wilson coefficients
$C_7^{\rm eff}$, $C_9^{\rm eff}$ and $C_{10}$.

Eq.~(1) can be used directly for inclusive $B$ decay. For $B\to
K^*\ell^+\ell^-$, hadronic matrix elements of quark bilinears give
well defined $B\to K^*$ form factors. Thus, even for exclusive
decay, the coefficients $C_9^{\rm eff}$, $C_{10}$ and $C_7^{\rm
eff}$ can be viewed as physical measurables, hence scheme and
scale independent, up to the definition of form factors.
Indeed, $C_7^{\rm eff}$ and $C_{10}$ in Eq.~(1) are at $m_B$
scale, with $C_7$ receiving large additive contributions from
other Wilson coefficients through operator mixing~\cite{pNLO},
\begin{equation}
C_7^{\rm eff} = \xi_7C_7 + \xi_8C_8 + \sum_{i=1}^6\xi_iC_i,
\end{equation}
where $\xi_i$ are QCD evolution factors.
However,
\begin{equation}
C_9^{\rm eff}(\hat{s}) = C_9 + {Y} (\hat{s}),
\end{equation}
is also a function of the dilepton mass through $Y(\hat{s})$, the
form of which can be found in Ref.~\cite{ABHH}, and depends on
long distance ($c\bar c$) effects.

Within SM, because of the near reality of $V_{ts}$ in the standard
phase convention, $C_7^{\rm eff}$, $C_9$ and $C_{10}$ are
practically real, with $C_9^{\rm eff}(\hat{s})$ receiving slight
complexity through $Y(\hat{s})$. A widely invoked~\cite{MFV}
``Minimal Flavor Violation" (MFV) scenario further asserts
(usually assuming the operator structure of SM) that there are no
further sources of flavor and $CP$ violation, other than what is
already present in SM. Indeed, many popular extensions of SM, such
as minimal supersymmetric SM \cite{bertolini} or two Higgs doublet
models \cite{weinberg}, tend to follow this pattern. With MFV as
the prevailing mindset, $C_7$, $C_9$ and $C_{10}$ are oftentimes
taken as real~\cite{LLST} tacitly, hence the focus only on
possible sign flips from large NP effects. For ${\cal A}_{\rm
FB}$, therefore, the main projection for the future has been the
sensitivity of the zero to NP~\cite{ABHH,GB}.

As a quantum amplitude, however, there is no reason {\it a priori}
why $C_7^{\rm eff}$, $C_9^{\rm eff}$ and $C_{10}$ in ${\cal
M}_{b\to s\ell^+\ell^-}$ 
should be real. Despite the suggested reality from SM and MFV,
whether they are real or complex should be {\it measured
experimentally}, and we will be able to do so in just a few years!
In fact, currently there are hints~\cite{HazBar} for ``anomalies"
in time-dependent and direct CP violation (CPV) measurements of
$b\to s\bar qq$ transitions. One possible explanation is NP in
$b\to s\bar qq$ electroweak penguins~\cite{HNS}, which are the
hadronic cousins of $b\to s\ell^+\ell^-$, but the latter is
clearly much less plagued by hadronic effects.

Motivated by possible hints for New Physics CPV in hadronic $b\to
s$ transitions, and in anticipation of major experimental progress
in near future, we explore how much ${\cal A}_{\rm FB}$ can differ
from SM by allowing associated effective couplings to be complex.
Constraints such as decay rates, of course, should be respected,
and one should check whether models exist where $C_7^{\rm eff}$,
$C_9^{\rm eff}$ and $C_{10}$ can be complex. We find, even without
enlarging the operator basis, from a theoretical standpoint, MFV
may be too strong an assumption.

Our insight comes as follows. Part of the impetus for MFV is the
good agreement between theory and experiment for inclusive $b\to
s\gamma$ rate, which provides a stringent constraint on NP.
However, while depending on the existence of a third generation
top quark, the $b\to s\gamma$ rate depends very little on the
precise value of $m_t$ when it is large. For $m_t$ in the range of
150 to 300 GeV, the $b\to s\gamma$ rate changes by only $\sim
30\%$.
In contrast, the $b\to s\ell^+\ell^-$ rate depends very
sensitively on $m_t$ through the effective $bsZ$ coupling, as we
stated from the beginning, changing by a factor of $\sim 4$
in the same $m_t$ range.

Suppose there are extra SM-like heavy quarks. These could be the
4th generation, or could be vector-like quarks that mix with the
top. Take the 4th generation as an example, the $b\to s\gamma$
rate is not sensitive to the existence of the $t'$ quark unless
$|V_{t's}^*V_{t'b}|$ is very large~\cite{HSS}. However, ``hard"
(sensitive to heavy quark mass) amplitudes such as $b\to
s\ell^+\ell^-$, $B_s$ mixing~\cite{HNSBsBsbar} etc. would be
easily affected by finite $V_{t's}^*V_{t'b}$, as $m_{t'} > m_t$ by
definition. Since $V_{t's}^*V_{t'b}$ should be in general
complex~\cite{AHphase}, so would $C_9^{\rm eff}$ and $C_{10}$ (and
$C_7^{\rm eff}$).
With this as an existence proof, we note further that the three
$[\bar sb]\,[\bar\ell\ell]$ terms in Eq.~(1) are 4-fermion
operators. The possible underlying New Physics is precisely what
we wish to probe at B factories and at the LHC.
Thus, despite the apparent success of MFV, we find the usual
assumption of near reality of $C_7^{\rm eff}$, $C_9^{\rm eff}$ and
$C_{10}$ unfounded. When sufficient data comes, the experimenters
are well advised to keep these parameters complex in doing their
fit.


We remark that, in fact within SM, $B\to \rho\ell^+\ell^-$ decay
exhibits partially the physics we talk about. A complex factor
$\delta_u \equiv ({V_{ud}^*V_{ub}})/({V_{td}^*V_{tb}})$ arises
from the $u$-quark current-current operator, as well as top quark
in the loop, making $C_9$ complex \cite{KS}. We will use this case
at the end as an illustration within SM.



In this study we shall keep the operator set as in SM, since
enlarging to include e.g. righthanded currents would not be
profitably probed in early years of LHC. In the same vein,
although inclusive $b\to s\ell^+\ell^-$ (and $b\to s\gamma$) is
theoretically cleaner, we focus on the experimentally more
accessible $B\to K^*\ell^+\ell^-$ (and $B\to K^*\gamma$). 
Experimental studies of inclusive processes usually apply cuts
that complicate theoretical correspondence.

Having allowed $C_7^{\rm eff}$, $C_9^{\rm eff}$ and $C_{10}$ to be
complex, we still need to consider the constraints. $C_7^{\rm
eff}$ is rather well constrained by $b\to s\gamma$ rate
measurement. We take a one sigma experimental range~\cite{PDG} for
$B\to K^*\gamma$ for our exclusive study. Likewise, inclusive
$b\to s\ell^+\ell^-$ measurement (by reconstructing a partial set
of $X_s$ states), as well as the exclusive $B\to K^*\ell^+\ell^-$,
provide constraints on $C_7^{\rm eff}$, $C_9^{\rm eff}$ and
$C_{10}$. At the moment, measurements are not precise enough, so
we use only the integrated rate for the exclusive channel, again
within one sigma experimental range. In the future, with high
statistics, one could use the differential $d{\cal B}/d{\hat s}$
rate, which is more powerful. We will plot $d{\cal B}/d{\hat s}$
as an illustration.

Our main focus is the ${\cal A}_{\rm FB}$ in exclusive $B\to
K^*\ell^+\ell^-$ decay. Assuming the form factors are real, we
have
\begin{eqnarray}
\frac{d {\cal A}_{\rm FB}}{d \hat s}
 &\propto&
      \Bigl\{ {\rm Re}\left(C_9^{\rm eff}C_{10}^*\right) VA_1
      \nonumber \\
 &+&  \frac{\hat m_b}{\hat s}\,{\rm Re}\left(C_7^{\rm eff}C_{10}^*\right)
       \bigl[ \left(VT_2\right)_- + \left(A_1T_1\right)_+ \bigr]
      \Bigr\}, \ \ \
\end{eqnarray}
where $(VT_2)_- = VT_2\,(1 - \hat m_{K^*})$, $(A_1T_1)_+ =
A_1T_1\,(1 + \hat m_{K^*})$, and $V$, $A_1$, $T_i$ are form
factors~\cite{ABHH}. We use the light-cone sum rule
(LCSR)~\cite{BZ} form factors in our numerical analysis. In
Eq.~(4) we have exhibited only the dependence on $C_9^{\rm eff}$,
$C_{10}$ and $C_7^{\rm eff}$, since it is customary to plot
$d\bar{\cal A}_{\rm FB}/d \hat s$ which is $d{\cal A}_{\rm FB}/d
\hat s$ normalized by the differential rate $d\Gamma/d\hat s$.
This reduces sensitivity to form factor models. The zero of
$\bar{\cal A}_{\rm FB}$ is often considered quite stable against
form factor variations~\cite{GB,ABHH}.

The Wilson coefficients are parameterized as,
\begin{eqnarray}
 C_7 &\rightarrow& C_{7}\,(1 + \Delta_7\,e^{i\phi_7}), \\ 
 C_9 &\rightarrow& C_{9}\,(1 + \Delta_9\,e^{i\phi_9}), \\ 
 C_{10} &\rightarrow& C_{10}\,(1 + \Delta_{10}\,e^{i\phi_{10}}),
\end{eqnarray}
with $\Delta_i = 0$ corresponding to SM. These Wilson coefficients
are evaluated at the electroweak scale, then evolved down to the
$m_B$ scale to be used in Eq.~(4). We do not include any
complexity from other Wilson coefficients. The tree level $C_1$
and $C_2$ are unchanged by NP, but as a simplifying assumption, we
ignore possible NP induced complexities through the gluonic
$C_{3{\rm -}6}$ and $C_8$ coefficients, which enter $C_7^{\rm
eff}$ and $C_9^{\rm eff}$ through operator mixing and long
distance effects (see Eqs. (2) and (3)). In practice, this should
not change our point.


Let us start with a SM-like framework, that is, viewing $B\to
K^*\ell^+\ell^-$ as induced by effective $bsZ$ and $bs\gamma$
couplings (and box diagrams). If such is the case, we expect $C_9$
and $C_{10}$ to be approximately the same, i.e.
\begin{equation}
 \Delta\equiv \Delta_9 \cong \Delta_{10},\ \ \ \ \phi\equiv \phi_9 \cong \phi_{10},
\end{equation}
in Eqs. (6) and (7), and one effectively has the parameters
$\Delta$, $\Delta_{7}$, $\phi$ and $\phi_7$, which covers the
usual case of wrong sign $C_7^{\rm eff}$. The 4th generation also
belongs to this scenario, with $V_{t's}^*V_{t'b}$ bringing in
complexity.

%
\begin{figure}[t!]
 \hbox{\hspace{0.03cm}
\hbox{\includegraphics[scale=0.415]{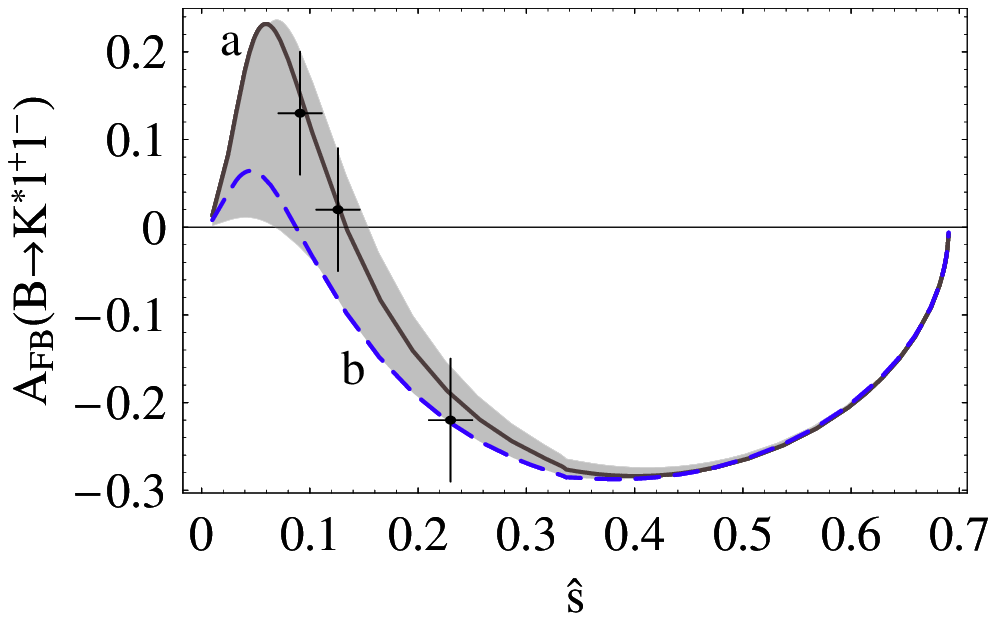}}
\hspace{-0.06cm}\hbox{\includegraphics[scale=0.415]{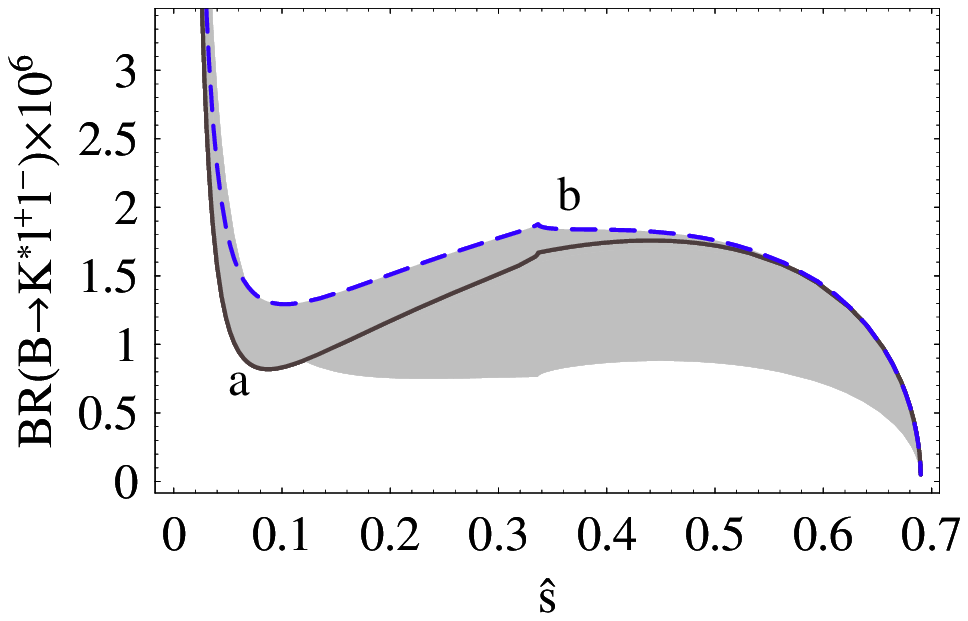}}}
 \caption{
 (a) $d\bar{\cal A}_{\rm FB}/d \hat s$ and (b) $d{\cal B}/d
 \hat s$ for $B\to K^* \ell^+\ell^-$. The shaded region is
 allowed by $C_9\cong C_{10}$, and Cases a (solid) and b (dash)
 are SM and 4th generation model, respectively.}
 \label{fig1}
\end{figure}


We plot $d\bar{\cal A}_{\rm FB}/d \hat s$ and $d{\cal B}/d{\hat
s}$ in Figs.~1(a) and (b), respectively, for SM and 4th generation
model (SM4). For SM4, we take the CKM parameters which yield the
correct $B_s$-$\bar{B}_s$ mixing~\cite{HNSBsBsbar}, predicts large
time-dependent CPV in $B_s$ decay, as well as
accommodating~\cite{HNS} the NP hints in CPV in $b\to s\bar qq$
decays.
We see that the zero of $d\bar{\cal A}_{\rm FB}/d \hat s$ has
shifted by a significant amount, with only a small positive value
below the zero. These are due to the enrichment of (mostly) the
$\phi$ phase. For larger $\hat s$, one has little difference in
$d\bar{\cal A}_{\rm FB}/d \hat s$ from SM, as the effect of
$C_7^{\rm eff}$ has damped away, while $C_9^{\rm eff}$ and
$C_{10}$ carry almost the same phase. The general appearance of
$d{\cal B}/d{\hat s}$ for SM and SM4 is very similar.

\begin{figure}[t!]
\hbox{\hspace{0.03cm}
\hbox{\includegraphics[scale=0.415]{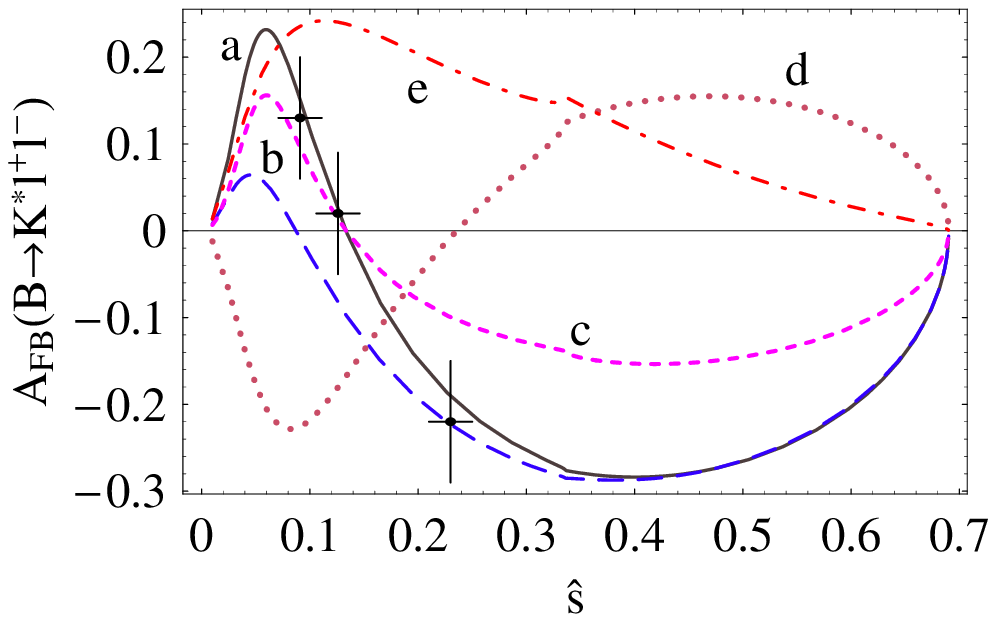}}
\hspace{-0.06cm}\hbox{\includegraphics[scale=0.415]{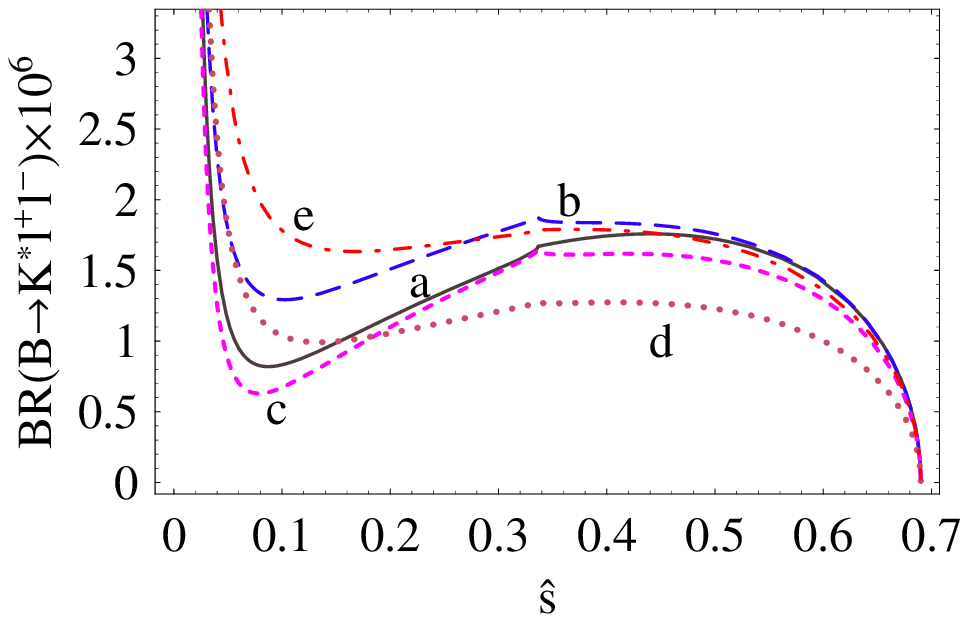}}}
\caption{
 (a) $d\bar{\cal A}_{\rm FB}/d \hat s$ and (b) $d{\cal B}/d \hat s$
 for $B\to K^* \ell^+\ell^-$ allowing all Wilson coefficients to be
 complex as in Eqs.~(5)--(7).}
 \label{fig2}
\end{figure}

A broader range is allowed by Eq.~(8). Keeping $B\to K^*\gamma$
and $K^*\ell^+\ell^-$ rates in $1\,\sigma$ experimental range and
exploring $\Delta$, $\Delta_{7}$, $\phi$ and $\phi_7$ parameter
space,
the results are plotted in Fig.~1 as the shaded area, which
illustrates the range of variation allowed by $C_{9}\simeq
C_{10}$.
This is just for illustration purpose and should not be taken as
precise boundaries. For instance, we see that below the SM zero
$d\bar{\cal A}_{\rm FB}/d \hat s$ could be very small, but the
shaded region for $d{\cal B}/d{\hat s}$ basically reflects the
$1\sigma$ constraints on $B\to K^*\gamma$ and $K^*\ell^+\ell^-$.
$d{\cal B}/d{\hat s}$ should also
be fitted in the future, 
but it depends directly on $B\to K^*$ form factors, especially the
overall scale.

We illustrate the power of early LHC data with the 2 fb$^{-1}$
study of LHCb, where $\sim 7700$ reconstructed $B\to
K^*\ell^+\ell^-$ events are expected. We take the simulated
errors~\cite{LHCb2fb} (with signal events generated according to
SM) for $d\bar{\cal A}_{\rm FB}/d \hat s$ from three bins, one
around the SM zero, one below and one above, and plot in Fig.~1(a)
to guide the eye. It should be clear that our suggestion can be
tested early on in the LHC era


The narrow, long ``tail" at $\hat s \gtrsim 0.3$ for $d\bar{\cal
A}_{\rm FB}/d \hat s$ indicates that Eq.~(8) is probably too
strong an assumption. Even if we keep the operator basis as in
Eq.~(1), treating these as 4-fermion interactions arising from
possible NP at short distance (for instance, $Z'$
models~\cite{BHI}), one should keep the full generality of
Eqs.~(5)--(7). We proceed to explore the parameter space as
before, keeping $B\to K^*\gamma$ and $K^*\ell^+\ell^-$ within
$1\,\sigma$ constraint. Indeed we find much richer possibilities
than Figs.~1(a) and (b). As plotted in Figs.~2(a) and (b), we
illustrate with the further cases of c, d and e. The SM and SM4
are Cases a and b, respectively, as in Fig.~1. The $\Delta_i$ and
$\phi_i$ values are given in Table I.

\begin{table}[t!]
\begin{center}
\begin{tabular}{ccccccccr}
\hline\hline
 \ Case \ & \ $\;\Delta_7$ \ & \ $\;\Delta_9$ \ & \ $\Delta_{10}$ \
      & $\phi_7$ \ & $\phi_9$ \ & $\phi_{10}$ \
\\ \hline
  b & $-0.2$ \ & $-0.9$ \ & $-0.9$ \
    & \ $65^{\circ}$ \ & $65^{\circ}$ \ & $65^{\circ}$ \\
  c & $-0.5$ \ & $1$ \ & $-0.5$ \ & \ $90^{\circ}$ \ & $270^{\circ}\;\,$ \ & $0$ \ \\
  d & $0$ \ & $-1.5$ \ & $-2.0$ \ & \ $0$ \ & $35^{\circ}$ \ & $0$ \ \\
  e & $-4.8$ \ & $-1.2$ \ & $-2.2$ \ & \ $0$ \ & $0$ \ & $0$ \ \\
\hline\hline
\end{tabular}
\caption{Parameter values for Cases b--e in Fig.~2, where Cases d
and e are already ruled out. The SM (Case~a) has $\Delta_i=0$. In
our numerics~\cite{pNLO}, we use $C_7^{\rm eff} \simeq 0.67\, C_7
- 0.18$, with $C_7 \simeq -0.20$, $C_9 \simeq 4.1$ and $C_{10}
\simeq -4.4$ at $M_W$ scale.}
 \label{tab1}
\end{center}
\end{table}

Case d has wrong sign $C_{10}$, while Case e has sign flip in both
$C_7^{\rm eff}$ and $C_{10}$ (equivalent to wrong sign $C_9$).
Both are already ruled out~\cite{AFBBelle} by Belle data. The
possibility of flipping only the sign of $C_7^{\rm eff}$ is ruled
out by rate constraints~\cite{GHM}, hence not plotted. Similar
scenarios have been considered in the literature, and we give
these cases to illustrate the versatility of Eqs.~(5)--(7). Though
ruled out, Case e is illuminating. From Table I, $\Delta_7 \sim
-4.8$ overwhelms the SM effect, flipping the sign of $C_7^{\rm
eff}$ (Eq.~(2)). At the same time, $C_{10}$ also flips sign, with
a much diminished $C_9$. However, even with no complex phases, the
large effects from Case e survives rate constraints, giving a more
pronounced low $q^2$ peak for differential rate, as seen in
Fig.~2(a), which lies outside of the boundary of the shaded region
in Fig.~1(a). This is because Eq.~(8) no longer holds. Similar
cases may exist that remain to be probed.

An interesting new scenario is illustrated by Case c, where
$d{\cal B}/d \hat s$ and the zero of $d\bar{\cal A}_{\rm FB}/d
\hat s$ are hard to distinguish from SM, but $d\bar{\cal A}_{\rm
FB}/d \hat s$ above the zero reaches only half the SM value.
Thus, a measurement of the zero does {\it not} pin down $C_9$.
The scenario can be tested
already with 1 ab$^{-1}$ data at B factories expected by 2008. If
such phenomena are discovered with, e.g. LHCb data, it would imply
NP that feed the $(\bar{s} \g_\mu L b)\,(\bar\ell\g^\mu\ell)$ and
$(\bar{s} \g_\mu L b)\,(\bar\ell\g^\mu\gamma_5\ell)$ operators
differently.

\begin{figure}[t!]
\begin{center}
\vskip 0.4cm \hskip 0.45cm
 \hbox{
\hbox{\includegraphics[scale=0.42]{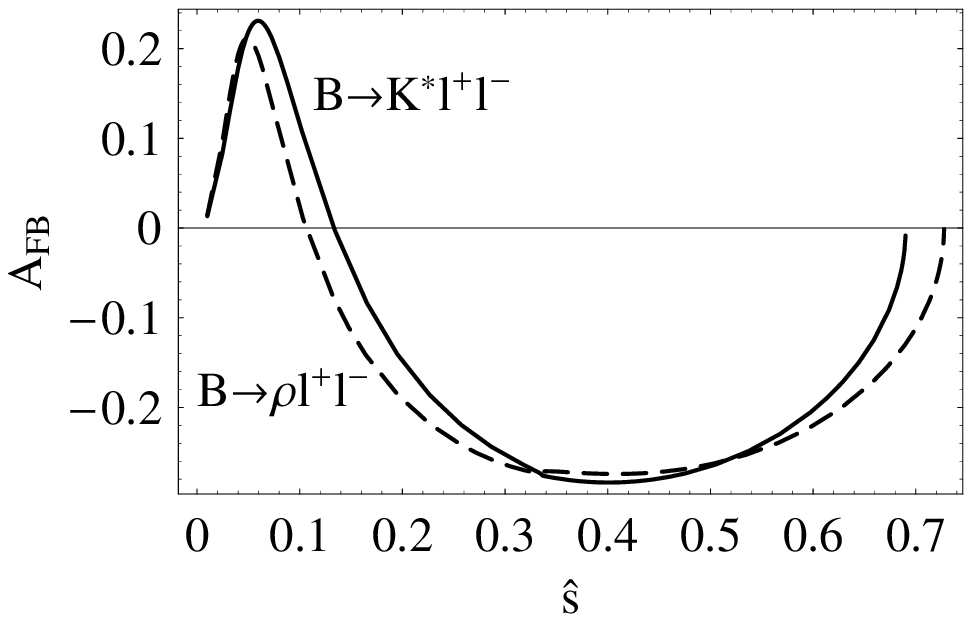}}}
 \caption{
 $d\bar{\cal A}_{\rm FB}/d \hat s$ for $B\to K^*\ell^+\ell^-$
 and $\rho\ell^+\ell^-$ within SM.}
 \label{fig3}
\end{center}
\end{figure}

We mark the simulated errors from the 2 fb$^{-1}$ study at LHCb as
before on Fig.~2(a), illustrating its power.
%
%
The actual possibilities are far richer. The shaded area of
Fig.~1(a) illustrates that, even with Eq.~(8) imposed, a broad
range is allowed for $\hat s < 0.2$. With the full freedom of
Eqs.~(5)--(7), the region allowed by rate constraint would likely
cover a large part of Fig.~2(a), which is up to experiment to
explore. One should keep the effective Wilson coefficients of
Eq.~(1) complex and use the general parametrization of
Eqs.~(5)--(7) to fit for $\Delta_i$ and $\phi_i$.
{\it Finite $\phi_i$ implies violation of MFV.}


Our suggestion of keeping the Wilson coefficients complex is not
just for NP. Even within SM, for the CKM suppressed decay $B\to
\rho\ell^+\ell^-$, one already expects~\cite{KS} complexity in
effective couplings, arising from both the $u$-quark and top
contributions. We plot $d\bar{\cal A}_{\rm FB}/d \hat s$ in Fig.~3
for $B\to K^*\ell^+\ell^-$ and $\rho\ell^+\ell^-$ in SM. Although
the difference is not so great, especially since with 2 fb$^{-1}$
data at LHCb one expects less than 200 $B\to\rho\ell^+\ell^-$
events with larger background, the different behavior in
$d\bar{\cal A}_{\rm FB}/d \hat s$ can be tested with a larger
dataset.



We offer some remarks before closing. We have focused on ${\cal
A}_{\rm FB}$ for exclusive $B\to K^*\ell^+\ell^-$, mostly because
of ease of experimental access, and with upgrade of statistics
imminent. Second, it is usually stressed that the zero of ${\cal
A}_{\rm FB}$ is insensitive to form factors. With NP sensitivities
now going beyond the zero, form factor issues would have to be
considered. One would have to combine the progress from form
factor models, lattice, as well as experimental studies of $B\to
\rho\ell\nu$.
One can of course try to measure ${\cal A}_{\rm FB}$ in inclusive
mode, and our discussion, starting with Eq.~(1), can be easily
employed and followed.
Third, while the 4th generation model provides a good example, our
approach is fully general and aimed at the experimental study, and
does not depend on specific models. In fact, the 4th generation
provides a good example against the prevailing MFV prejudice that
limits the perspective for flavor and $CP$ violation expectations
in $b\to s$ transitions. It is our opinion that the study of $b\to
s$ transitions is still in its infancy, and is the least
constrained. Imposing MFV may be overstretching our experience
from other areas of flavor violation. It is up to experiment to
reveal what may be in store for us in the excellent probe of
${\cal A}_{\rm FB}$.
%
%
Finally, the 1~ab$^{-1}$ final data at B factories would only give
limited improvement on the existing result. The next round of
major improvement would come from LHC. The Super B factory upgrade
in the future could bring back competitiveness of $e^+e^-$
machines, especially for inclusive studies.

In summary, we have explored the $CP$ conserving consequences of
complex Wilson coefficients on the forward-backward asymmetry
${\cal A}_{\rm FB}$ in $B\to K^*\ell^+\ell^-$ decay. The
possibilities are much broader than the usual consideration of
sign flips under minimal flavor violation framework. In view of
hints of CP violation anomalies in $b\to sq\bar q$ decays, the
large increase in statistics with the advent of LHC would make
${\cal A}_{\rm FB}$ one of the cleanest probes for New Physics in
the near future.

\vskip 0.3cm \noindent 
 We thank A.~Buras, B.~Grinstein, P.~Koppenburg, Z.~Ligeti, and I.~Stewart
 for discussions.
 This work is supported in part by NSC 95-2119-M-002-037, NSC 95-2811-M-002-031,
 and NCTS of Taiwan.


\begin{thebibliography}{99}
%
%
\bibitem{HWS}
  W.S.~Hou, R.S.~Willey and A.~Soni,
  Phys.\ Rev.\ Lett.\  {\bf 58}, 1608 (1987).
%
\bibitem{Houmumu}
  G.W.S.~Hou,
  Int.\ J.\ Mod.\ Phys.\ A {\bf 2}, 1245 (1987).
%
\bibitem{AFB}
 A.~Ali, T.~Mannel and T.~Morozumi,
 Phys.\ Lett. \ B {\bf 273}, 505 (1991).
%
\bibitem{AFBBelle}
 A.~Ishikawa {\it et al.} [Belle Collab.],
  Phys.\ Rev.\ Lett.\  {\bf 96}, 251801 (2006).
%
%
\bibitem{GHM}
 P.~Gambino, U.~Haisch and M.~Misiak,
  Phys.\ Rev.\ Lett.\  {\bf 94}, 061803 (2005).
%
\bibitem{LHCb2fb}
 J. Dickens [LHCb Collab.], talk
 at CKM 2006 Workshop, Nagoya, Japan, December 2006.
%
\bibitem{ABHH}
 A.~Ali, P.~Ball, L.T.~Handoko and G.~Hiller,
  Phys.\ Rev.\ D {\bf 61}, 074024 (2000).
%
\bibitem{pNLO}
  We use partial NLO for our numerics, where some small terms,
  such as a mild $s$ dependence of $C_7^{\rm eff}$,
  are dropped. Parameters are as in
  H.M.~Asatrian, K.~Bieri, C.~Greub, A.~Hovhannisyan,
  Phys.\ Rev.\ D {\bf 66}, 094013 (2002).
%
\bibitem{MFV} See e.g.
  A.J.~Buras,
  Acta Phys.\ Polon.\ B {\bf 34}, 5615 (2003);
  B. Grinstein, talk presented at CKM 2006 Workshop, Nagoya,
  Japan, December 2006;
  and references therein.

%
\bibitem{bertolini}
See e.g. S.~Bertolini, F.~Borzumati, A.~Masiero and G.~Ridolfi,
Nucl.\ Phys.\ B {\bf 353}, 591 (1991).
%
\bibitem{weinberg}
 S.L.~Glashow and S.~Weinberg,
 Phys.\ Rev.\ D {\bf 15}, 1958 (1977),
 using the term ``Natural Flavor Conservation".
%
\bibitem{LLST}
  See e.g.
  K.S.M.~Lee, Z.~Ligeti, I.W.~Stewart and F.J.~Tackmann,
  hep-ph/0612156.
  These authors emphasize optimized $q^2$ binnings for early test
  of SM.
  %
\bibitem{GB}
G.~Burdman,
Phys.\ Rev.\ D {\bf 57}, 4254 (1998);
M.~Beneke and T.~Feldman,
Nucl.\ Phys.\ B {\bf 612}, 25 (2001).
%
\bibitem{HazBar}
  M. Hazumi and R. Barlow, plenary talks at 33rd International Conference
  on High Energy Physics, Moscow, Russia, July 27 - August 2, 2006.
%
\bibitem{HNS}
  W.S. Hou, M. Nagashima and A. Soddu,
  Phys.\ Rev.\ Lett.\ {\bf 95}, 141601 (2005);
  W.S.~Hou, H.n. Li, S.~Mishima and M.~Nagashima,
  hep-ph/0611107.
%
\bibitem{HSS}
  W.S.~Hou, A.~Soni and H. Steger,
  Phys.\ Lett.\ B {\bf 192}, 441 (1987).
%
\bibitem{HNSBsBsbar}
  W.S.~Hou, M.~Nagashima and A.~Soddu,
  hep-ph/0610385.
%
\bibitem{AHphase}
  A. Arhrib and W.S. Hou,
  Eur. Phys. J. C {\bf 27}, 555 (2003).
%
\bibitem{KS}
 F.~Kr\"uger and L.M.~Sehgal,
 Phys.\ Rev.\ D {\bf 56}, 5452 (1997).


\bibitem{PDG}
 W.M. Yao {\it et al.} [Particle Data Group], J. Phys. G {\bf 33},
 1 (2006).
%
\bibitem{BZ}
P.~Ball and R.~Zwicky,
Phys.\ Rev.\ D {\bf 71}, 014029 (2005).
%
\bibitem{BHI}
  G.~Buchalla, G.~Hiller and G.~Isidori,
  Phys.\ Rev.\ D {\bf 63}, 014015 (2001).
  Starting with a framework similar to ours, these authors
  quickly focus on real Wilson coefficients.

\end{thebibliography}
\end{document}